\documentclass[twocolumn,showpacs,aps,prb,amsmath,amssymb,floatfix]{revtex4}
\usepackage{graphicx}
\usepackage{epstopdf}
\usepackage{amssymb}
\usepackage{dcolumn}
\usepackage{bm}
\begin{document}
\title{Vacancy mediated magnetization and healing of a graphene monolayer}
\author{E. Nahkmedov$^{1,2}$, E. Nadimi$^{3}$, S. Vedaei$^{3}$, O. Alekperov$^{2}$, F. Tatardar$^{2,4}$, A. I. Najafov$^{2}$, I. I. Abbasov$^{5}$, 
and A. M. Saletskii$^{6}$ }
\affiliation{ 
$^{1}$ Faculty of Physics, Moscow State University, Baku branch, str. Universitetskaya 1, AZ-1144 Baku, Azerbaijan,\\
$^2$Institute of Physics, Azerbaijan National Academy of Sciences, H. Cavid ave. 33, AZ1143 Baku, Azerbaijan,\\
$^{3}$ Center for Computational Micro and Nanoelectronics (CCMN), Faculty of Electrical Engineering, 
K. N. Toosi University of Technology, Tehran, Iran\\
$^{4}$Khazar University, Mahsati str. 41, AZ 1096, Baku, Azerbaijan,\\
$^{5}$ Azerbaijan State Oil and Industry University, Azadlig ave. 20, Baku, Azerbaijan\\
$^{6}$ Faculty of Physics, Leninskie Gory 1-2, 119991 Moscow State University, Moscow, Russian Federation
}
\date{\today}
\begin{abstract}
Vacancy-induced magnetization of a graphene layer is investigated by means of a first principle DFT 
method. Calculations of the formation energy and the magnetization  by creating the different 
number of vacancies in a supercell show that a clustering with big number of vacancies in the cluster is rather
favorable than that of isolated vacancies, homogeneously distributed in the layer. The magnetic moment 
of a cluster with big number of vacancies is shown to be not proportional with the vacancy concentration, which is in
good agreement with the recent experimental results. Our studies support the idea that although the vacancies 
in a graphene create a magnetic moment, they do not produce a magnetic ordering.  It is shown that although the Lieb's
rule for the magnetization in a hexagonal structure violates, two-vacancies, including a di-vacancy, in the supercell 
generate quasi-localized state when they belong to the different sublattices, and instead two-vacancies
generate an extended state when they belong to the same sublattices. Analytical investigation of the dynamics of 
carbon atom- and  vacancy-concentrations according to the non-linear continuity equations shows that the vacancies, 
produced by irradiation at the middle of a graphene layer, migrate to the edge of the sample resulting in a
specific {\it 'segregation'} of the vacancy concentration and self-healing of the graphene.   
\end{abstract}
\maketitle
\section{Introduction} 
Investigation of graphene, the single two-dimensional  (2D) sheet of graphite, is one of the priority 
direction among of allotropic modifications of carbon and related nanostructures due to its topological 
properties. Unusual properties of pristine graphene such as ballistic electron propagation with extremely 
high carrier mobility of $\mu_e=10^4~cm^2V^{-1}s^{-1}$ at room temperature \cite{ngm04, geim09}, existence 
of massless, chiral low-energy excitations, characteristic to Dirac fermions, observation of anomalous
integer quantum Hall effect (IQHE) even at room temperature \cite{njzm07}, makes it an attractive
material for electro-technical industry. Recent experimental and theoretical investigations show that 
a graphene sheet with vacancies heal themself under thermal annealing or electric potential spanning  
\cite{wrgg09, zzly15, zbrn12, rzbb12, sl01, zrbn12, cscx13, glin10, wycd12, spb13, bpag16, llly14, ogc13}, 
so that the vacancies either migrate toward the crystal edge or form a big hole near extended defects 
like grain boundary or dislocation in the crystal. This property of the graphene may have a great impact 
to the graphene based electronic technology.  

Defects engineering may enormously change kinetic and magnetic characteristics of graphene, extending 
a scope of its application in spintronics.
The defects are usually produced as point defects, like substitutional- and interstitial- impurities and 
vacancies, either in the fabrication processes of a sample  or by means of external factors, such as 
radiation or heavy-ion bombardment and chemical doping of a sample (see, e.g. [\onlinecite{atd12}] for a review).
The vacancies are formed in a crystal by reconstruction of the lattice  as a result of knock-on of carbon
atoms from the graphene lattice or/and as a result of bond rotations, e.g. the formation of Stone-Wales
(55-77) and 8-ring defects, under bombardment with high-energy particles. A beam of the high-energy particles acts 
as a local heat source and thus helps to overcome the defect formation energy barrier.
Note that a new class of the carbon-based materials 
has been experimentally revealed \cite{mmkr01}, 
which has a nanoporous structure and it is catalycally active. The nanoporous carbon is less 
ordered than graphite but not completely amorphous. Theoretical studies have shown \cite{cs06} 
that an introduction non-hexagonal rings by means of vacancies into carbon structure most likely 
method to get a nanoporous carbon structure. 

Although many effects observed in graphene are attributed to the presence of vacancies and defects, 
we will focus here particularly on a magnetic response and self-repairing of graphene in the presence 
of vacancies.

Magnetic response associated with vacancies produced by the irradiation of graphene with high-energy
protons and carbon ($C^{4+}$) ions has been studied recently as experimentally \cite{ubgg10, whsz09, ckf09} as well 
as theoretically \cite{yh07,yazyev10, mlfn04, ztkv07, bkk11, py12, cwtd15} by means of he molecular dynamics 
simulations and DFT based {\it ab-initio} methods.
Many experimental observations reported in the literature provide inconsistent even contradictory results.
Experimental evidence of ferromagnetic order has been observed \cite{whsz09} in irradiated graphene,
the origin of which was suggested by the authors to be the defects located in the structure.
Magnetic ordering was mainly observed in graphitic materials \cite{ckf09}, 
which was explained by the existence of the localized electronic states at grain boundaries of highly
oriented defective pyrolytic graphite. Some observation of ferromagnetism even anti-ferromagnetism 
seems to be artifacts.

{\it Ab-initio} density-functional theory (DFT) investigations of magnetization produced by vacancies 
in a graphene are mainly performed \cite{yazyev10, mlfn04, cs06, ztkv07, bkk11, py12, cwtd15, czll11} 
by a superlattice method, where a vacancy is created in a supercell of the big number of unit cells, 
periodically extended with images of the original vacancy. Spin orientations in each supercell become
the same, yielding seemingly a ferromagnetic signature of total magnetization even in the absence of a spin-spin 
interactions. Nevertheless, we believe that a magnetic
ordering in a system should be achieved by exchange interactions between the magnetic moments.  
A numeric computation reveals that a vacancy in graphene \cite{psn08, pgsp06, ubgg10} introduces 
a semi-localized $\pi$-midgap state with Coulomb-like $\sim 1/r$ decay potential. The single-atom defects 
create in the graphene a quasi-localized state at the Fermi level \cite{rgs00, mf89, rmg05, pgsp06}.
The graphene structure can be viewed as two interpenetrating hexagonal sublattices of carbon atoms 
labeled as $A$ and $B$ ones, forming a bipartite lattice.  A defect introduced into the $A$
sublattice results in a quasi-localized state due to the $p_z$ orbitals of carbon atoms in the 
$B$ sublattice and vice versa. A short-range repulsive Hubbard interactions between vacancies 
at higher concentrations and with half-filled band may result in a ground state with total spin
$S=|N_A-N_B|/2$ according to Lieb's theorem \cite{lieb89}, where $N_A$ and $N_B$ are the number of 
vacancies in $A$ and $B$ sublattices, correspondingly, of a bipartite lattice. 
The balance between $N_A$ and $N_B$ is destroyed at e.g. zigzag edges, 
which may provide a low-temperature background magnetism observed in a graphene sample \cite{skmt18}.   

Recent investigation of a graphene sheet by controllable doping  it with fluorine adatoms and creation 
of vacancies by irradiation \cite{snrn10, nstl12, ntsl13} has indicated that both defects induce 
a magnetic moments with spin $1/2$. Nevertheless, 
they do not result in a magnetic ordering, rather an induced paramagnetism down to liquid helium 
temperatures was observed in the doped samples. Low temperature measurements of pristine graphene 
by using  SQUID magnetometry disclose strong diamagnetism
\cite{snrn10}, and show a tiny background paramagnetism, observable at $T<50~K$. The samples under 
investigation were shown \cite{snrn10, nstl12} to consist of electronically decoupled 
$10-50~nm$ mono- and bilayer graphene crystallites, which are aligned parallel to each other.  Evaluation of the spin
number $N$ from the measurement data shows that the number of paramagnetic center is proportional to
the defect concentration $x$ for a small concentration ($x<0.5$), and the dependence is more complicated
for higher concentrations. For each concentration $x$, the measured number of paramagnetic centers is three
orders of magnitude less than the measured number of defects in the samples. This experimental fact can be 
understood such that only one of out $\sim 1000$ defects contributes to the paramagnetism in contradictory
to other experiments as well as ab-initio investigations \cite{yazyev10, mlfn04, cs06, ztkv07, py12, cwtd15} 
that each defect contributes one Borh magneton $\mu_B$ to the total magnetization. Note that the vacancy magnetism 
in graphene was recently shown \cite{ntsl13} to be originated from two approximately equal contributions: 
one from the dangling bonds and other from itinerant magnetism. 

High resolution transmission electron microscope (HRTEM) monitoring  of a graphene sheet at $80 keV$ operating 
energy has displayed \cite{wrgg09} a formation of a hole, which seems to support the above described experiment
\cite{nstl12} on vacancy magnetism. Indeed, as we will show, agglomeration of the vacancies in a big hole results in
not a linear increase of the magnetic moment with vacancy concentration. Theoretical investigations of time-evolution 
of a graphene with dense vacancies by means of non-equilibrium molecular dynamics \cite{zzly15} reveal also
a tendency toward a formation of the haeckelites in the case with small number of the vacancies, while 
forming a large hole as the number of vacancies increases. 

A systematic scanning transmission electron microscopy (STEM) study of a suspended graphene layer  
\cite{zbrn12, rzbb12}, deliberately introduced vacancies and deposited by metal atoms such as Cr, Ti, Pd, 
Ni, Al except for gold, shows that nanoscale holes were etched in the structure due to interactions of metal 
atoms with graphene. The nanoholes were observed at room temperature in ultrahigh vacuum ($6 \times 10^{-9}~mbar$) 
under low-energy $60~keV$ (lower than the threshold 'knock-on' energy $86~meV$ for a carbon atom \cite{sl01} 
in graphene) that electron-beam scanning acts as catalysts for etching holes on the graphene surface. 
Instead, the mending and filling of many-vacancy holes (over 100 vacancies) was observed \cite{zrbn12} 
in room-temperature metal-catalyzed etching STEM experiments under the same conditions described above, 
provided a reservoir of loose carbon atoms is readily available nearby holes. This process was interpreted 
by the authors as a dislodging of carbon adatoms from the graphene
surface by the scanning electron-beam and dragging them to the edge of the holes, which
results in  refilling the holes by random combination of 5, 6, 7, and 8 carbon atom rings.

Healing effect in graphene, where the vacancies were produced by employing plasma bombardment 
\cite{cscx13}, has been performed by thermal annealing without external carbon atoms source 
in the temperature interval starting from $300^{\circ}C$ up to $900^{\circ}C$. For higher temperatures
the self-repairing was shown to stop. According to the results of Raman, x-ray photoemission spectroscopy (XPS), 
$HRTEM$, and electrical transport measurements the healing takes place by annihilation of displaced carbon atoms 
with vacancies with assistance of thermal energy. Healing was shown to become more difficult when the 
size of the vacancies' hole increases. Formation of monovacancy defects in the finite graphene flakes \cite{glin10} 
and graphene nanoribbons \cite{wycd12} was shown to be size-dependent, and that the vacancy defects migrate 
toward the edge at higher temperatures, as a result of which the structures heal themselves. DFT calculations 
and molecular dynamics simulations of the vacancy migration in graphene flakes show \cite{spb13}
thermally activated motion of vacancy toward the edge  occurs even at room temperature whereas
the probability of return motion back to the middle is negligible.

Our {\it ab-initio} investigation of the vacancies and analytically study of their migration in a graphene layer supports 
the experimental results on magnetism measured in \cite{nstl12} as well as the experiments on self-healing of a graphene. 
Analysis of our calculation results shows that the formation energy and magnetization of divacancy is lower than those 
of single vacancies.  Single vacancies are enough mobile, and they diffuse toward the extended defects like a grain boundary, 
a dislocation or a sample edge (which can be considered as an extended defect) or coalesce with other vacancies, forming 
poly-vacancies with lower energy and magnetic moment. We show that a merging of a single vacancy with  polyvacancies, resulting 
in defect cluster or hole with bigger size, is energetically favorable. According to our calculations the formation energy 
of a even-fold vacancy like di- or tetra-vacancy is lower than old-fold vacancies. Therefore, they are more stable.  
Magnetic moments of the clusters with minimal formation energies are lower too. The magnetic moment of the graphene 
is determined by edge structure of the vacancy clusters. So that only that defect on the e.g. A sublattice that has 
no counterpart on the B sublattice will contribute to the magnetization. The authors of Ref. [\onlinecite{wycd12}]  
considered a hole of $N>1$ multiple vacancies, concluded that the dangling bonds these structural defect is 
proportional to the circumference of the hole or to $\sqrt{N}$, and therefore its formation energy is also 
proportional to $\sqrt{N}$. In contrast we show that a multiple vacancy hole is structured in different 
modifications, and at least one of them possesses minimal dangling bonds and, consequently minimal magnetic 
moment. Thus, there is a tendency of clustering of vacancies into big holes, instead of homogeneous distribution 
of isolated vacancies over the system in the irradiation process. It is worthy to note that total energy calculations 
have been done in Ref. [\onlinecite{czll11}] for nanoholes of various sizes, containing  up to 60 vacancies, in a big 
supercell with 288 carbon atoms in graphene by applying ab-initio DFT method. Although the number of possible holes 
for a given number vacancies $N$ increases sharply with $N$, it is hard task to calculate all possible cluster structures 
with higher N. Nevertheless, the results of  Ref. [\onlinecite{czll11}] yield extremely useful information on stability 
and magnetism of holes with a great number of vacancies.

Such kind of ``{\it segregation}'' seems to be a result of two-dimensional character 
of the graphene sheet where an off-diagonal long range order (ODLRO) is absent. The crystalline structure of 
the graphene is controlled by power-like order. Correlations of the vacancies seem to  be managed also with 
power-like order, which compete with correlations between the carbon atoms in the structure. In order to understand 
the vacancy hole formation and healing mechanism in a graphene layer, a phenomenological kinetic model is employed by us 
for a migration of carbon atoms through vacancies. This model has been implemented to numerical studies of a segregation 
problem under irradiation of binary alloys \cite{anthony72, ohl73, oh72, oswt74, sth73, ow74, jl76, jl77, ropw78, jl78, wol79}, 
where $A$- and $B$-type atoms diffuse over vacancies and interstitials. We simplified the segregation problem for 
diffusion of carbon atoms over vacancies in one-dimensional ($1D$) case, and solved analytically the non-linear 
differential equations for carbon atoms- and vacancies- concentrations. The obtained results show that a vacancy 
created, say in the middle of the sample, diffuses to the edge of the sample. We think that boundary of samples can be 
considered as an `` extended defect'' like dislocations, grain boundaries or a hole of a big size. 
Therefore, the vacancies diffuse and are incorporated, commonly say around extended defects in the sample.        

The paper is structured as follow: the computational method, employed for our {\it ab-initio} calculations is 
described in the next Section II. Section III presents the obtained results of {\it ab-initio} DFT calculations. 
Section IV provides our analytic investigation of diffusion-segregation problem for carbon atoms through 
vacancies, and migration of the vacancies toward the sample boundary.  Our conclusions and speculations are 
given in Section V.

\section{Computational methods}
SIESTA code is employed \cite{oas96, sagg02} in our spin-polarized DFT-based {\it ab-initio} 
calculations, where the standard double-$\xi$ basis with polarization orbitals ($DZP$) is used. 
The generalized-gradient approximation (GGA) is utilized to calculate the exchange correlation 
term \cite{pbe96}. The interaction between the valence electrons and the atomic core is taken 
into account by using standard norm conserving Troullier-Martins pseudopotential. The unit cell 
of a pristine mono-layer graphene is initially relaxed and then three different supercells  
of $50$, $98$, and $162$ atoms are constructed. The real space integration grid had $200~Ry$ 
cutoff  and $50~meV$ energy shift. Spin resolved calculations are performed in most cases. 
$k$-point sampling of the Brillouin zone was performed by using the Monkhorst-Pack method 
\cite{mp76}. In order to get an optimal self-consistency convergence, the density of 
electronic states (DOS) was calculated for different $k$-point meshing. 
\begin{table}[t]
\caption{Formation energies of mono-, di- and tri-vacancy for the supercells with 50, 98, and
162 atoms.}
\begin{tabular}{c  c  c  c} 
\hline \hline
\textbf{Formation energy} &  \textbf{1}       & \textbf{2}         &  \textbf{3}\\ 
\textbf{(eV)}             & \textbf{vacancy}  & \textbf{vacancies} & \textbf{vacancies}  \\ [0.5ex]  
\hline
Supercell with 50 atoms         & 7.03     & 5.78  & 8.50  \\ 
Supercell with 98 atoms         & 6.99     & 5.76    & 7.74 \\
Supercell with 162 atoms        & 6.78      & 5.21   & 7.52  \\ [1ex] 
\hline \hline
\end{tabular}
\label{table1}
\end{table}
Our calculations show that $5 \times 5 \times 1$ $k$-point mesh is a good and optimal choice to 
have a balanced accuracy and computation time. To obtain the equilibrium geometry we relaxed all 
the atoms after creation each vacancy until the forces acting on them were smaller than $0.01~eV/\AA$. 
The formation energy for mono-, di- and tri-vacancy are presented in Table \ref{table1} for the 
supercells with 50, 98, and 162 carbon atoms. The formation energy of 
the vacancies  changes slightly with the supercell size (see, Table \ref{table1}) as a result of 
interaction of the defects with their images in the neighboring auxiliary cells (we used periodic 
boundary conditions in all space directions). 
Our calculations of the formation energy, DOS, and the 
magnetization show that a good convergence can be reached for a supercell with at least 98 atoms.
Therefore, our results presented below were done for a supercell with 98 atoms.     

\section{Distribution of the vacancies and their magnetizaton}
\begin{figure}
\begin{tabular}{cc}
\resizebox{.48\textwidth}{!}{%
\includegraphics[width=1cm]{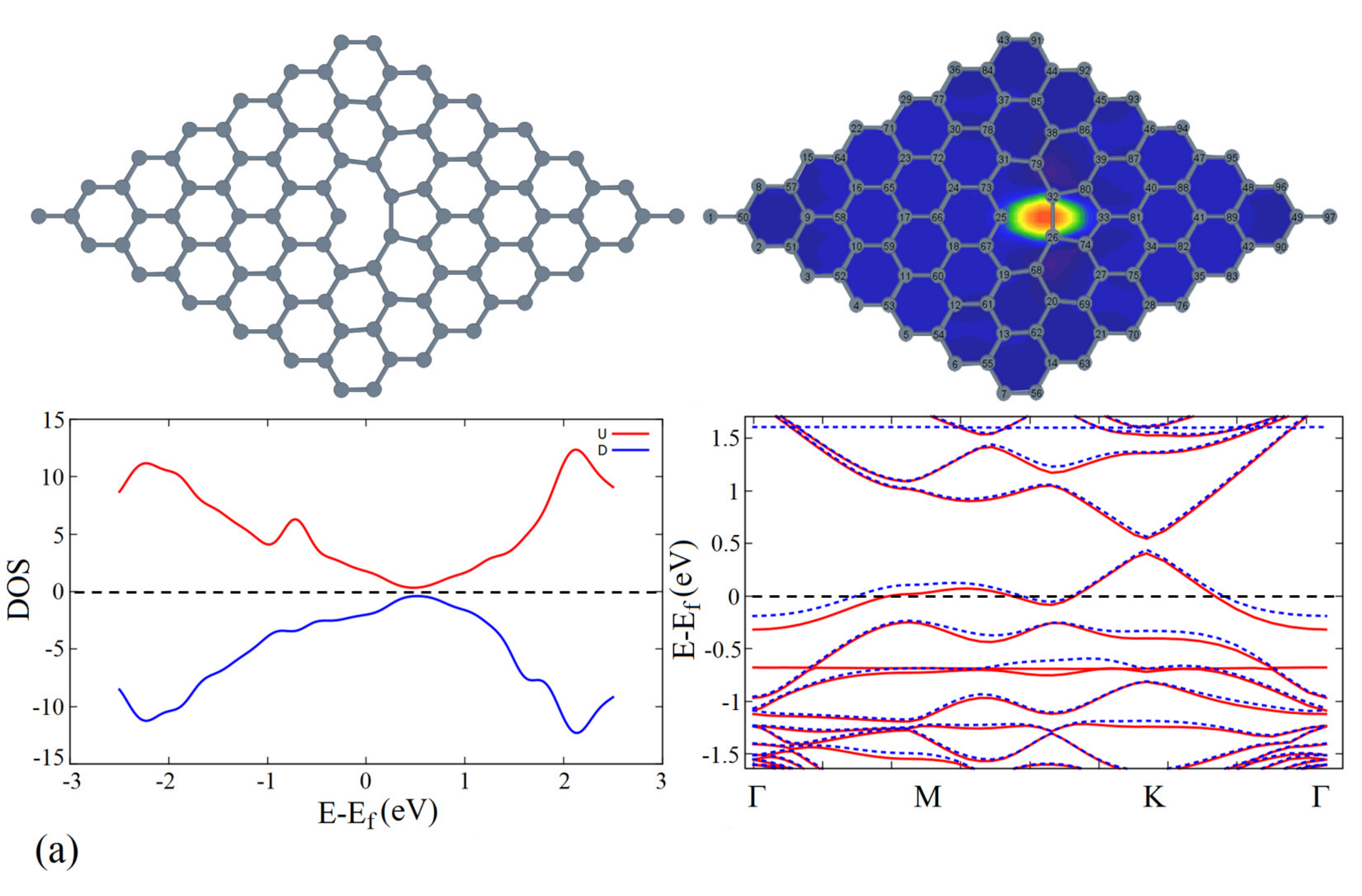}}\\
\resizebox{.48\textwidth}{!}{%
\includegraphics[width=1cm]{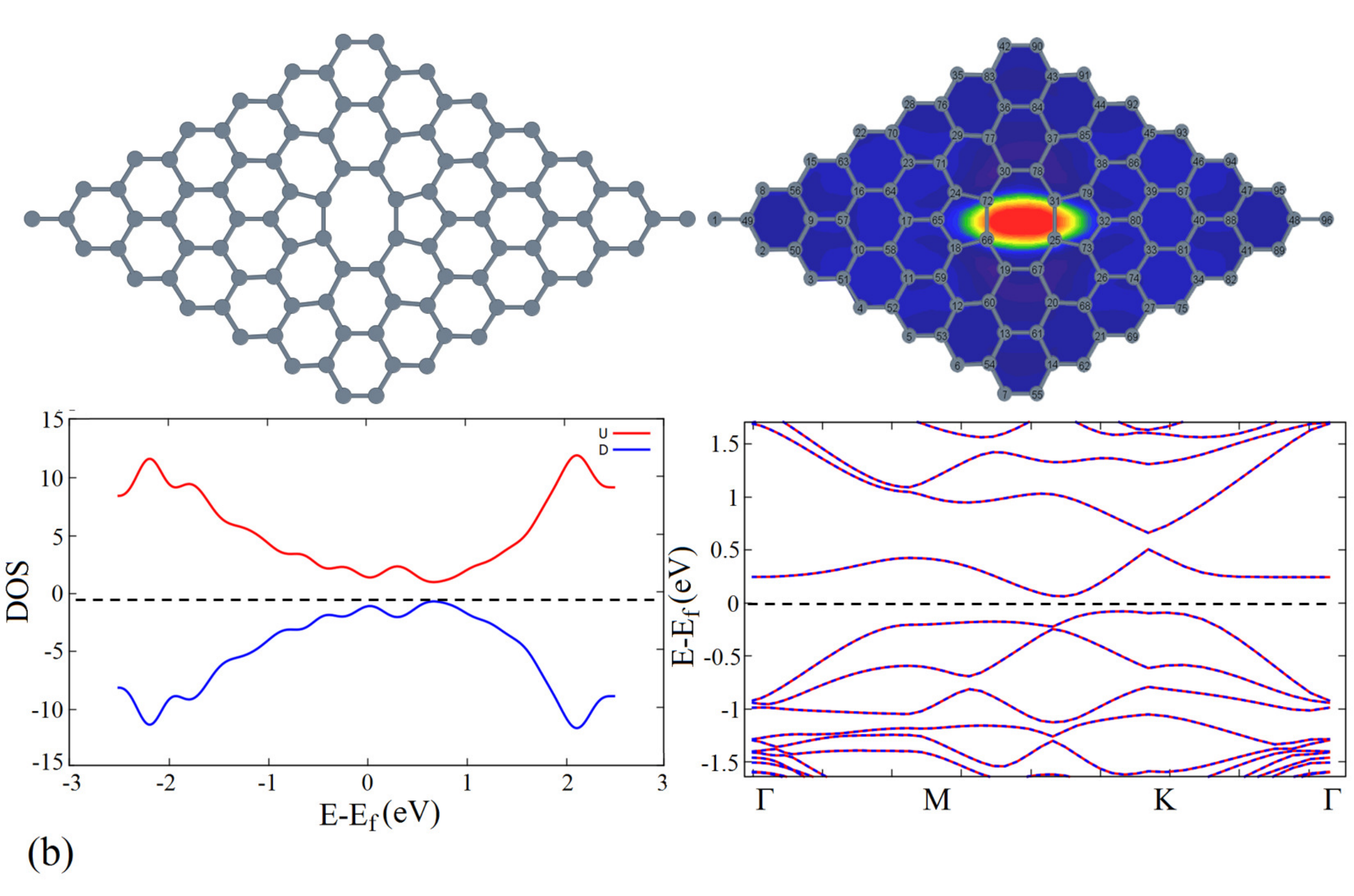}}\\
\resizebox{.48\textwidth}{!}{%
\includegraphics[width=1cm]{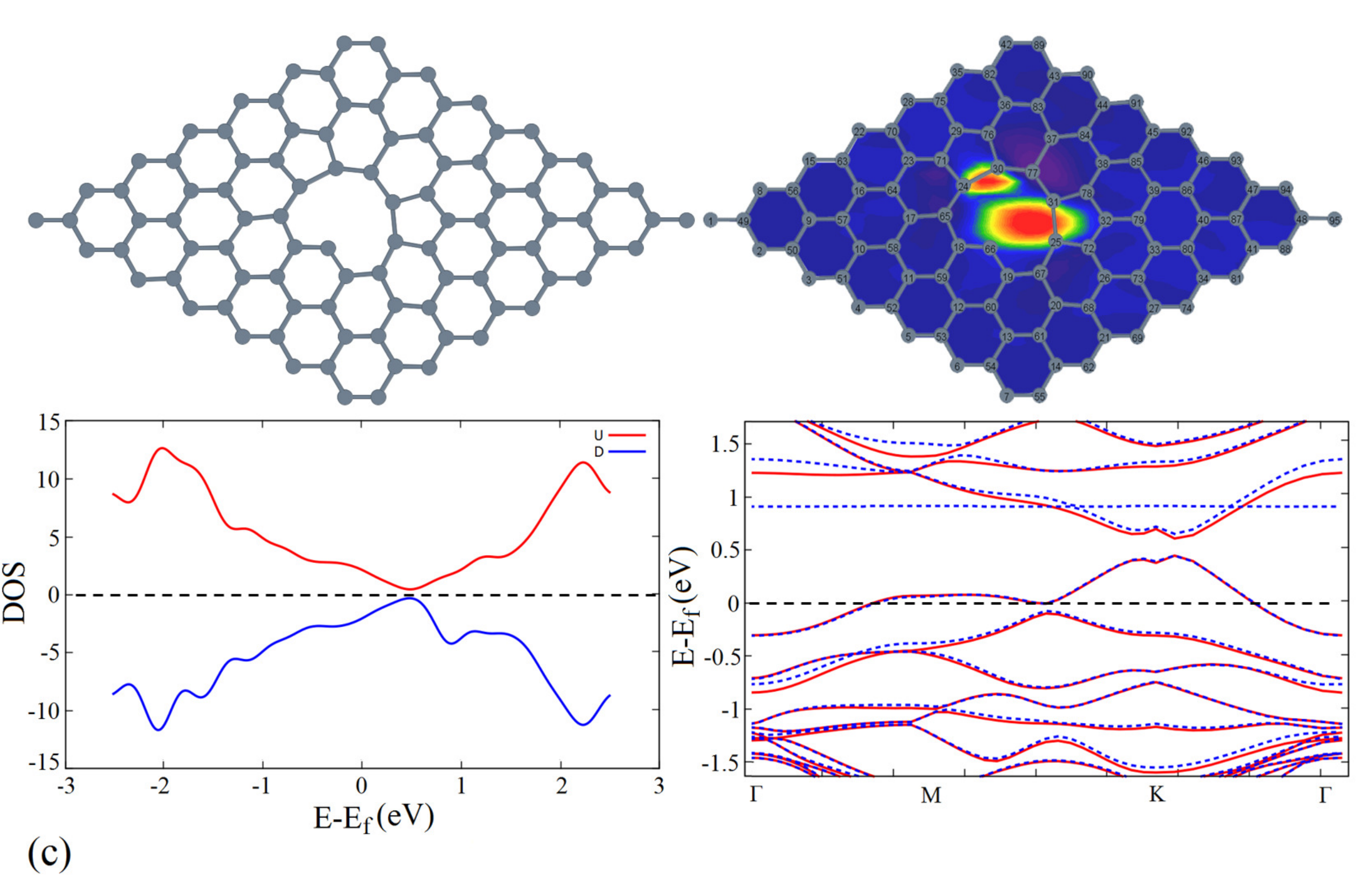}}
\end{tabular}
\caption {The relaxed structure and  the deformation potentials around the vacancies, the spin-polarized DOS 
and the band structures for (a) mono-, (b) di-, and (c) tri-vacancy, when one carbon atom or two and three 
nearest-neighboring carbon atoms are removed in the unit cell.  The DOS, corresponding to spin-up and spin-down 
states, are plotted respectively by red- and blue-curves in opposite directions for clarity.}
\label{Vac1}
\end{figure}

\begin{figure}[h]
\begin{tabular}{cc}
\resizebox{.48\textwidth}{!}{%
\includegraphics[width=1.0cm]{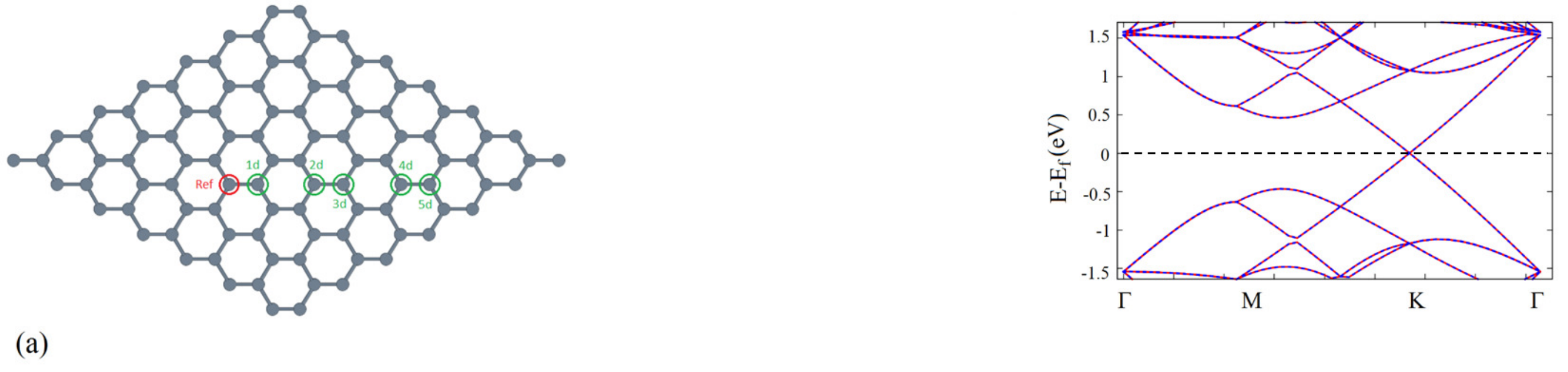}}\\
\resizebox{.48\textwidth}{!}{%
\includegraphics[width=1.0cm]{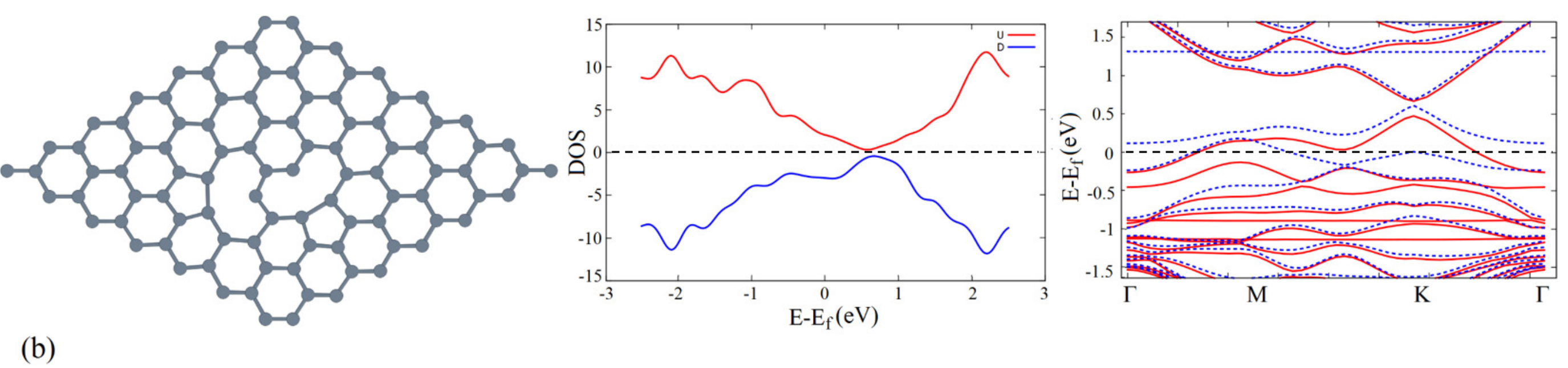}}\\
\resizebox{.48\textwidth}{!}{%
\includegraphics[width=1.0cm]{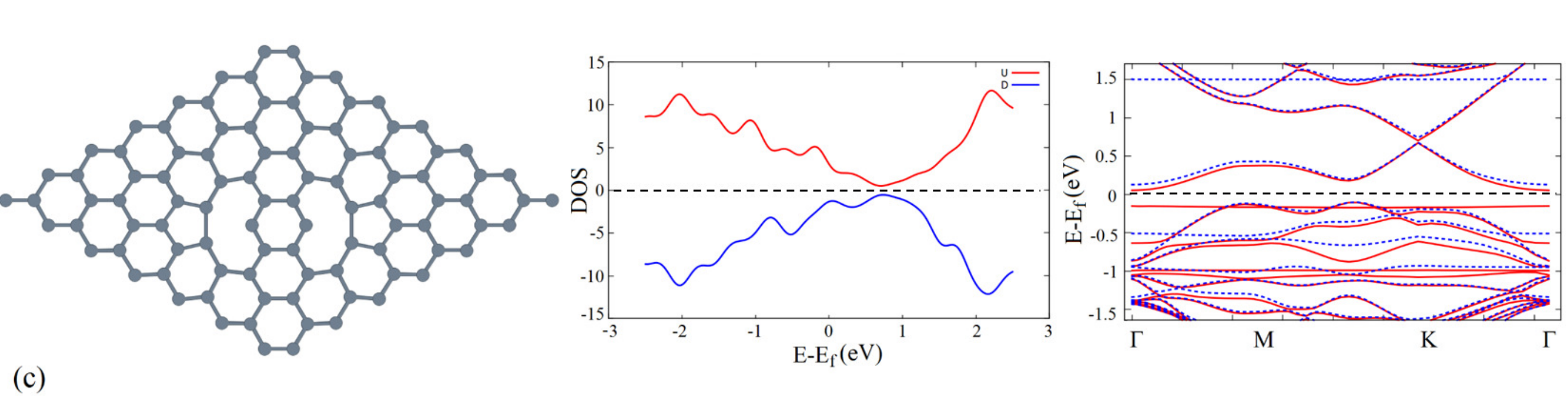}}\\
\resizebox{.48\textwidth}{!}{%
\includegraphics[width=1.0cm]{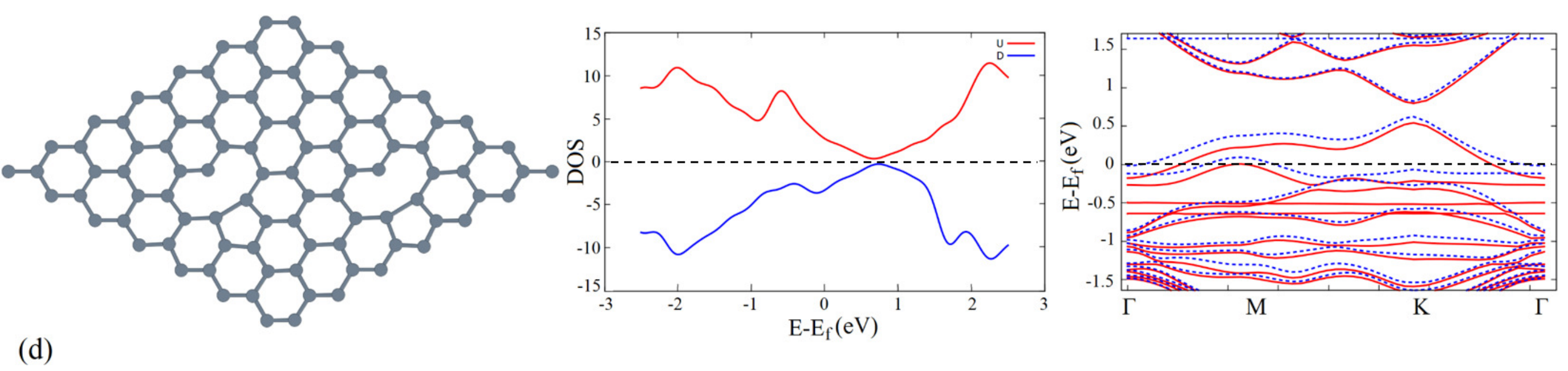}}\\
\resizebox{.48\textwidth}{!}{%
\includegraphics[width=1.0cm]{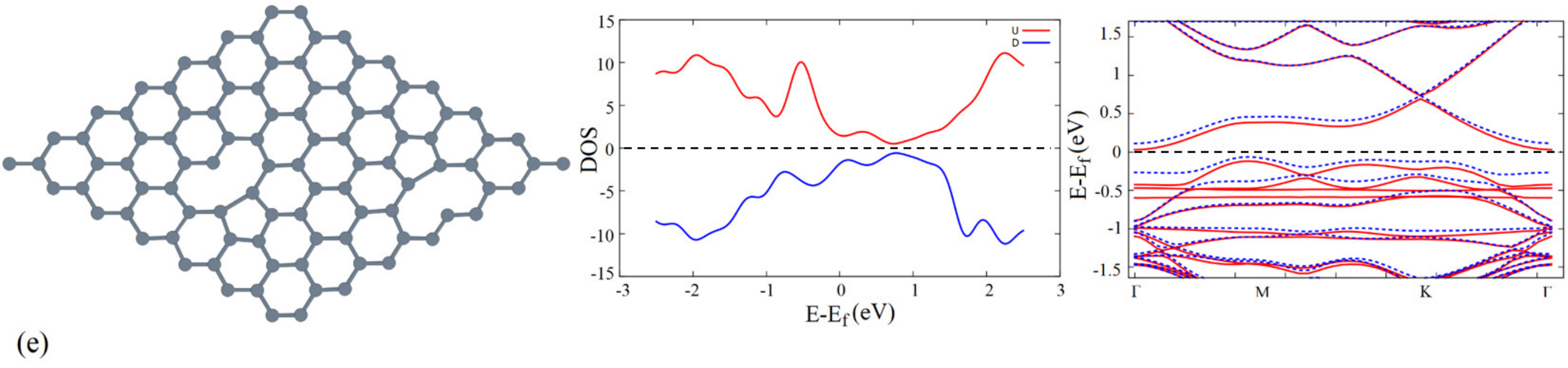}}
\end{tabular}
\caption {(a) Band structure of a pristine graphene monolayer, and a supercell of two-vacancy states, produced 
by removing a reference (red circle) and second $C$-atoms 1, 2, 3, and 4 (green circles), located correspondingly 
in the 1d, 2d, 3d, and 4d interatomic distances with d being equal to $C-C$ bond length. The  relaxed supercell, 
DOS, and the band-structure of two vacancies located at (b) one atomic distance (1d), (c) 2d-, (d) 3d-, and  
(e) 4d-interatomic distance.}
\label{Vac2}
\end{figure}

Our aim in this investigation is to understand a development of  the magnetization  with 
increasing the vacancy concentration and a restructuring tendency due to the vacancies 
migration in the graphene mono-layer. We calculated for this purpose the formation energy, the band 
structure, and the DOS of a supercell of 98 atoms with up to four vacancies in different 
configurations. The {\it C-C} bond length of a pristine graphene lattice is calculated
to be $1.409 ~\AA$, $1.412~\AA$, and $1.412~\AA$ for the supercells correspondingly with 50, 98, and 162 atoms, 
which is in good consistent with  the experimental value of $1.42~\AA$.
\begin{table}[t]
\caption{Formation energy and total spin polarization for the supercell of 98 atoms with different vacancies.}
\begin{tabular}{c  c  c} 
\hline \hline
\textbf{Type}  &\textbf{Formation energy}  &\textbf{Spin polarization} \\ 
     {}          & \textbf{(eV)}    & \textbf{ $(Q_{up}-Q_{down})$} \\ [0.5ex]
\hline
Single defect                     & 6.99     & 1.13 \\ 
DV 5-8-5                          & 5.76     &  0.0 \\ 
DV 555-888          & 4.52       &0.0 \\
3 vac. 55-10       &  8.84    & 1.00 \\
3 vac. 999-3        & 17.49       & 4.00\\
3 vac. 555-11       & 11.91   & 1.54\\
4 vac. 555-9        & 7.91     &  0.0 \\
4 vac. 5-12-5a      & 11.27    & 2.02 \\
4 vac. 5-12-5b      & 11.27     & 2.02 \\
4 vac. 55-12         & 13.33     & 1.99 \\
4 vac. 12-55        & 13.33      & 1.99 \\ [1ex] 
\hline \hline
\end{tabular}
\label{table2}
\end{table}
Creation of one vacancy, producing pentagon-nonagon type {\it 5-9} defect in the supercell, results in 
the appearance of three dangling $\sigma$- bonds, 
two of which rebind again with each other due to the Jahn-Teller distortion around vacancy. 
Jahn-Teller distortion deforms the lattice around the vacancy and breaks the threefold symmetry; 
the geometrical distortion is relaxed off behind the vacancies. Covalent-bond coupling between 
two dangling bonds  of the second-nearest-neighboring atoms belonging to  the same sublattice 
stabilizes the vacancy extended states. The third dangling bond is left unsaturated and contributes 
$\approx 1~\mu_B$ in magnitude magnetic moment to the intrinsic magnetization. Removing one-, two- and
three-carbon atoms located in the nearest-neighboring sites in the unit cell results in formation of 
correspondingly {\it 5-9}, {\it 5-8-5}, and {\it 55-10} type defective configurations in the supercell 
(see, Fig. \ref{Vac1}). DOS, the band structure, and the distributions of the deformation 
potential around the mono-, di-, and tri-vacancy distortions are depicted in Fig. \ref{Vac1}. 
The deformation potential is seen to be significant within the unit cell around the vacancy. 
Therefore, interaction of vacancies through the distortion potential seems to be like to the 
dipole interaction and it would be important only for vacancies placed  nearest-neighbor each other. 
The vacancies located far away each other may be considered practically as isolated vacancies. 
Formation energy of a single vacancy is calculated to be $6.99~eV$ (see, Table \ref{table2}).
This result is slightly lower than those obtained by other authors $E_f \approx 7.5~eV$ 
\cite{eteh03, mlfn04, klfn06}, nevertheless it is in good consistence with the experimental data 
of $7.0 \pm 0.5~eV$ \cite{tm78}.
The spin-polarized electronic band structures of a defective graphene with one vacancy and a pristine graphene 
are shown in Fig. \ref{Vac1}a and Fig. \ref{Vac2}a, correspondingly, for comparative study. 
The band gap of the perfect graphene is closed at the border of the Brillouin zone, yielding two Dirac 
cones. A single vacancy seems to remove the band crossing degeneracy at $K$-points of the Brillouin zone, 
and introduces at the same time two extended defect levels corresponding to spin-up and spin-down states 
at  the Fermi level. Assuming that a continuous irradiation of a sample 
will create isolated vacancies, the magnetization of the sample would monotonically increase. 
Furthermore, in the absence of spin-spin correlations between localized dangling bonds, the magnetic 
moments of the vacancies belonging to different sublattice will partially compensate each other and 
reduce the total magnetization. Nevertheless, the magnetization under this assumption would monotonically 
increase with vacancy concentration. Our calculations below show that creation of mono vacancies 
monotonically distributed throughout the structure is not energetically favorable, and there is a tendency 
to form a cluster or hole of vacancies by coalescing them each other or collapsing around 
the extended defects like grain boundaries, dislocations or the sample boundary.

A divacancy (DV) is produced by the removal of two nearest-neighboring carbon atoms, composing the 
so-called {\it 5-8-5} defect of an octagon and a pair  of pentagons as it is shown in Fig. \ref{Vac1}b. 
We calculated the formation energy of DV, presented in Table \ref{table2}, which is equal to 
$\epsilon^f_{DV}\approx 5.76~eV$. {\it Ab-initio} calculations of the DV formation energy by other 
authors \cite{eteh03, klfn06} indicate $\sim 8eV$. The energy $\epsilon^f_ {DV}$ is smaller than that 
for a monovacancy in a graphene, confirming that DV formation in the graphene 
is rather favorable. Divacancies create a quasi-localized state with energies close to the Fermi energy $E_F$; 
at the same time they remove the band crossing, destroying the gapless states at {\it K}-points of the Brillouin 
zone (see, Fig.\ref{Vac1}b). All the dangling bonds corresponding to the same sublattice rebind again, 
and therefore the DV does not generate a magnetization (see, Table \ref{table2}). 

The {\it 5-8-5} divacancy, consisting of two pentagons and one octagon, has another modification {\it 555-777}, 
which is structured by three pentagons and three heptagons. The {\it 555-777} defect 
is formed by removing two carbon atom and by additional creation of a Stone-Wales (SW) defect \cite{sw86}, e.g. 
by additional rotation of one {\it C-C} bond in the octagon. 
Although {\it 555-777} DV, with formation energy smaller then that for mono-vacancy (see, Table \ref{table1}),  
is suggested \cite{kiyl11} to be more stable, {\it 5-8-5} DV is responsible  for migration of DV. 
{\it Ab-initio} calculations in \cite{kiyl11} show that subsequent transformations of {\it 555-777} DV into
{\it 5-8-5} then {\it 5-7-7-5} and again {\it 5-8-5} DV create finally {\it 555-777} DV, which differs from the initial
one by rotation. {\it 5-8-5} DV transforms to two sequential {\it 5-7-7-5} metastable DV structures, which differ 
each other by rotation of heptagons and shifting of pentagons. The metastable {\it 5-7-7-5} DV structure finally 
turns to stable {\it 5-8-5} DV, the position of which  differs slightly from the position of initial {\it 5-8-5} DV. 

In order to see clearly the differences between a divacancy and other possible two-vacancy structures 
the latter are created,  as it is seen from Fig. \ref{Vac2}, in four different positions within 
the supercell, belonging both to the same or to different sublattices. Note that study of two-vacancies 
problem is instructive one in order to understand Lieb's rule and contribution of the dangling bonds to 
the magnetization.   

Two removal vacancies' positions in the supercell is depicted in Fig. \ref{Vac2}a, where the vacancies at 
'even distance' and the reference vacancy, showing by red, belong to the same sublattice (see, Figs. \ref{Vac2}b 
and $d$), instead of the 'odd distance' vacancy and the reference vacancy belong to the different sublattices 
(see, Figs. \ref{Vac2}c and $e$). According to the Lieb's rule, the magnetization in the first case should be 
finite, whereas in the second case it should be zero. The formation energies and the magnetizations of 
different two-vacancy topologies are presented in Table \ref{table3}. As it is seen from Fig. \ref{Vac2},  
all relaxed supercell structures contain two pentagon-nonagon {\it 5-9} defects, each of which is typical of a 
mono-vacancy in Fig. \ref{Vac1}a, even when two vacancies are located in $1d$ distance. Two vacancies in 
larger distance do not interact practically each other, and the formation energies as well as the 
magnetizations become approximately two times higher than that of a mono-vacancy.
\begin{figure}[t]
\begin{tabular}{cc}
\resizebox{.48\textwidth}{!}{%
\includegraphics[width=1cm]{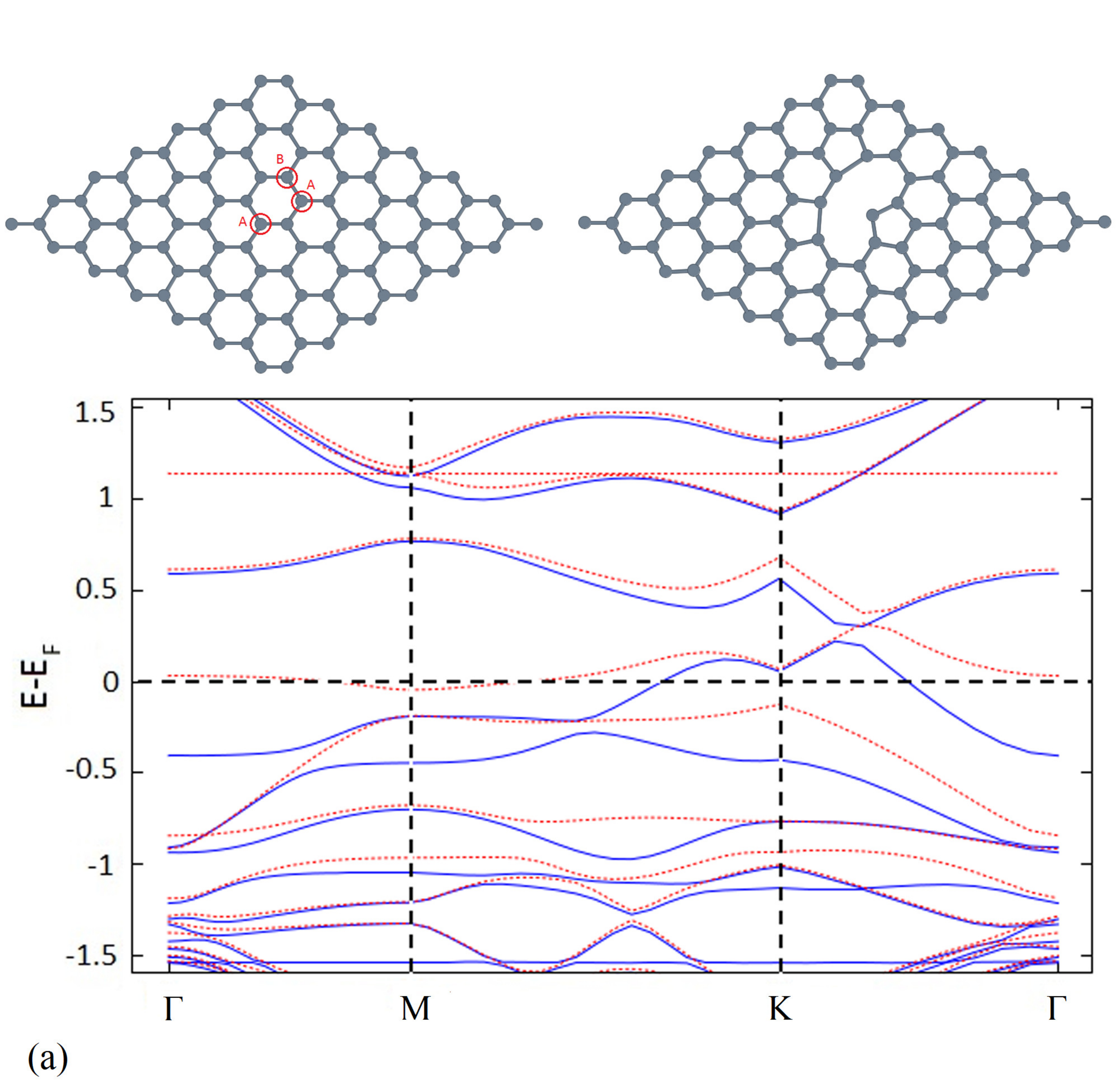}}\\
\resizebox{.48\textwidth}{!}{%
\includegraphics[width=1cm]{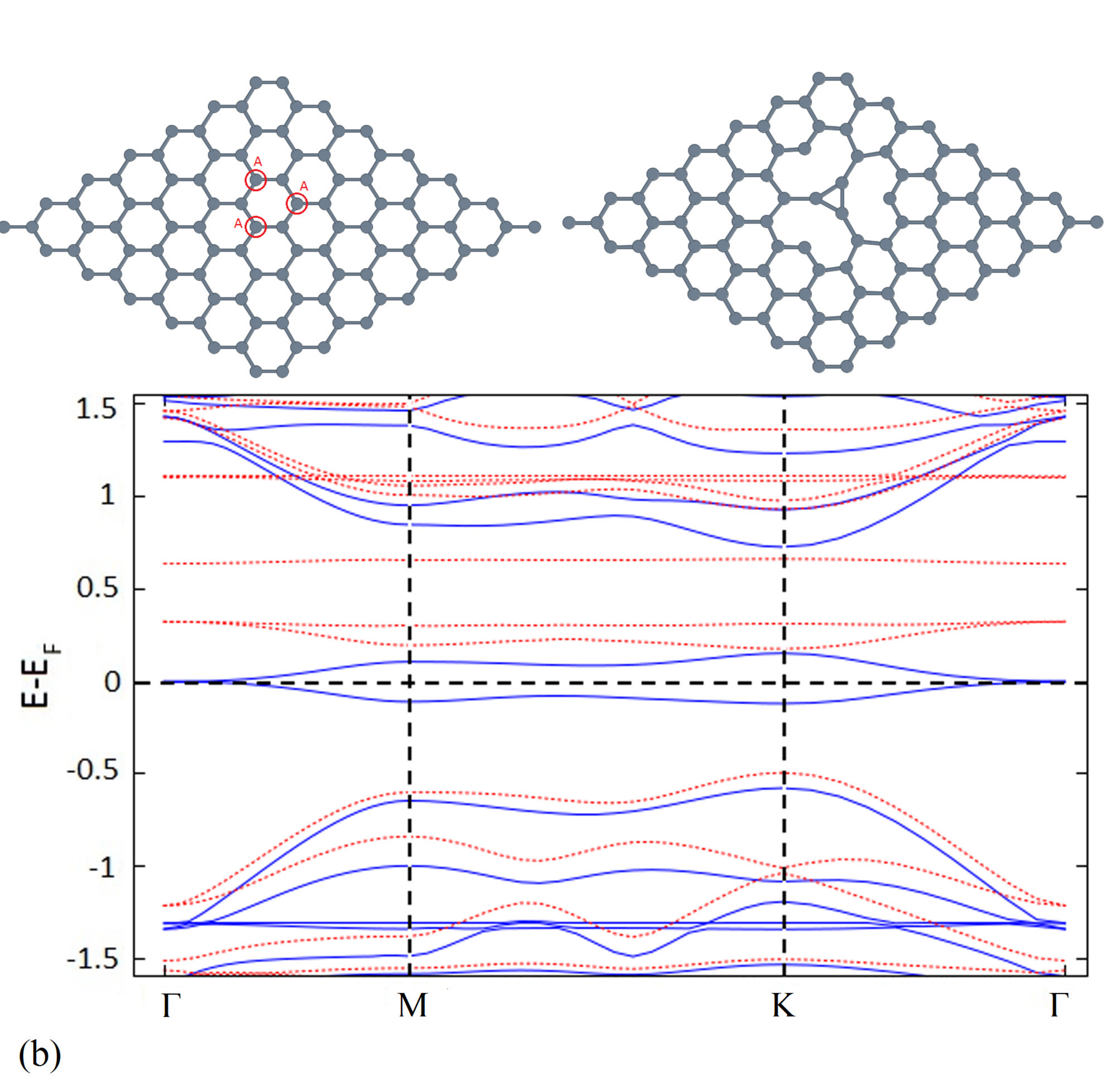}}
\end{tabular}
\caption {Two other possible structures for three-vacancy defect, where not all three vacancies are bounded 
each others in difference from the tri-vacancy structure in Fig. \ref{Vac1}c with three bounded removal $C$-atoms.}
\label{Vac3}
\end{figure}
All the two-vacancy structures (in Fig. \ref{Vac2}) except a di-vacancy in Fig. \ref{Vac1}b have dangling bonds. 
Consequently, the local DOS of DV for spin-up and spin-down states, shown  in Fig. \ref{Vac1}b correspondingly by 
red (above zero) and blue (below zero) curves, coincides completely at each points. The local DOS corresponding to 
two opposite spin polarizations for all other vacancy configurations, presented in Figs. \ref{Vac2}, differ each 
other due to existence of the dangling bonds. The magnetization of all these configurations becomes non-zero 
irrespective to which sublattice the vacancies belong. This fact show that the Lieb's rule seems to violate for a 
graphene mono-layer, and the magnetic moment of the supercell with several vacancies is determined with the number 
of the dangling bonds but not with the difference of the number of atoms in each sublattices. The band structures 
of di-vacancy and all other two-vacancy states are presented correspondingly in Fig. \ref{Vac1}b and 
Figs. \ref{Vac2}b, c, d, e. It is necessary to pay attention to the fact that two-vacancies (including the DV) 
generate quasi-localized state (see, the band structures in Fig. \ref{Vac1}b and Figs. \ref{Vac2}c, e) when 
they belong to the different sublattices, and instead two-vacancies generate an extended state (see the band 
structures in Figs. \ref{Vac2}b, d) when they belong to the same sublattices. Furthermore, the band structures, 
corresponding to spin-up and spin-down states, coincide each other only in the presence of the di-vacancy 
due to absence of a dangling bond. All other two-vacancy structures contain dangling bonds,
resulting in a splitting of opposite spin-polarized states in the band structures. 
  
\begin{figure}
\begin{tabular}{cc}
\resizebox{.45\textwidth}{!}{%
\includegraphics[width=1cm]{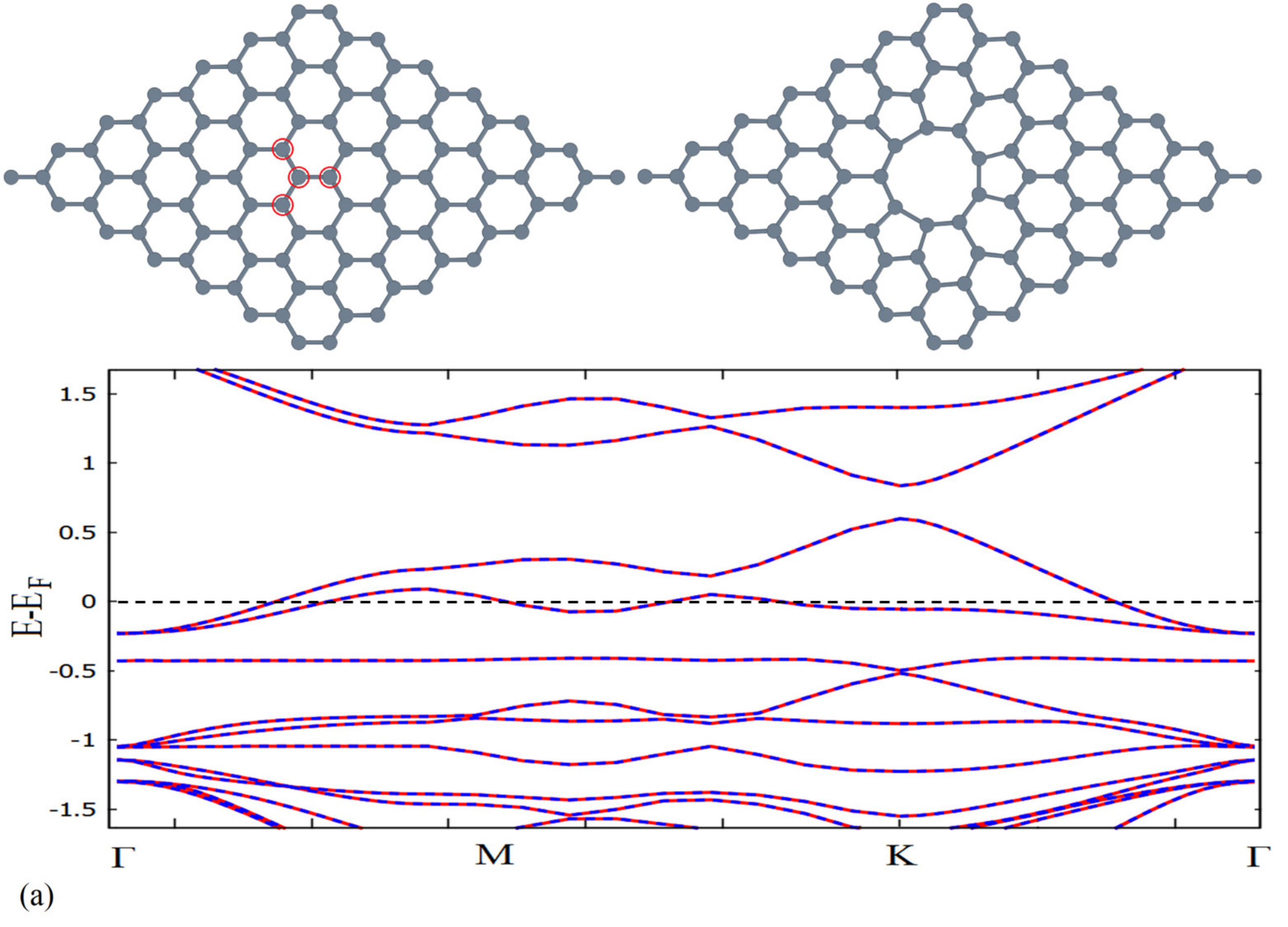}}\\
\resizebox{.45\textwidth}{!}{%
\includegraphics[width=1cm]{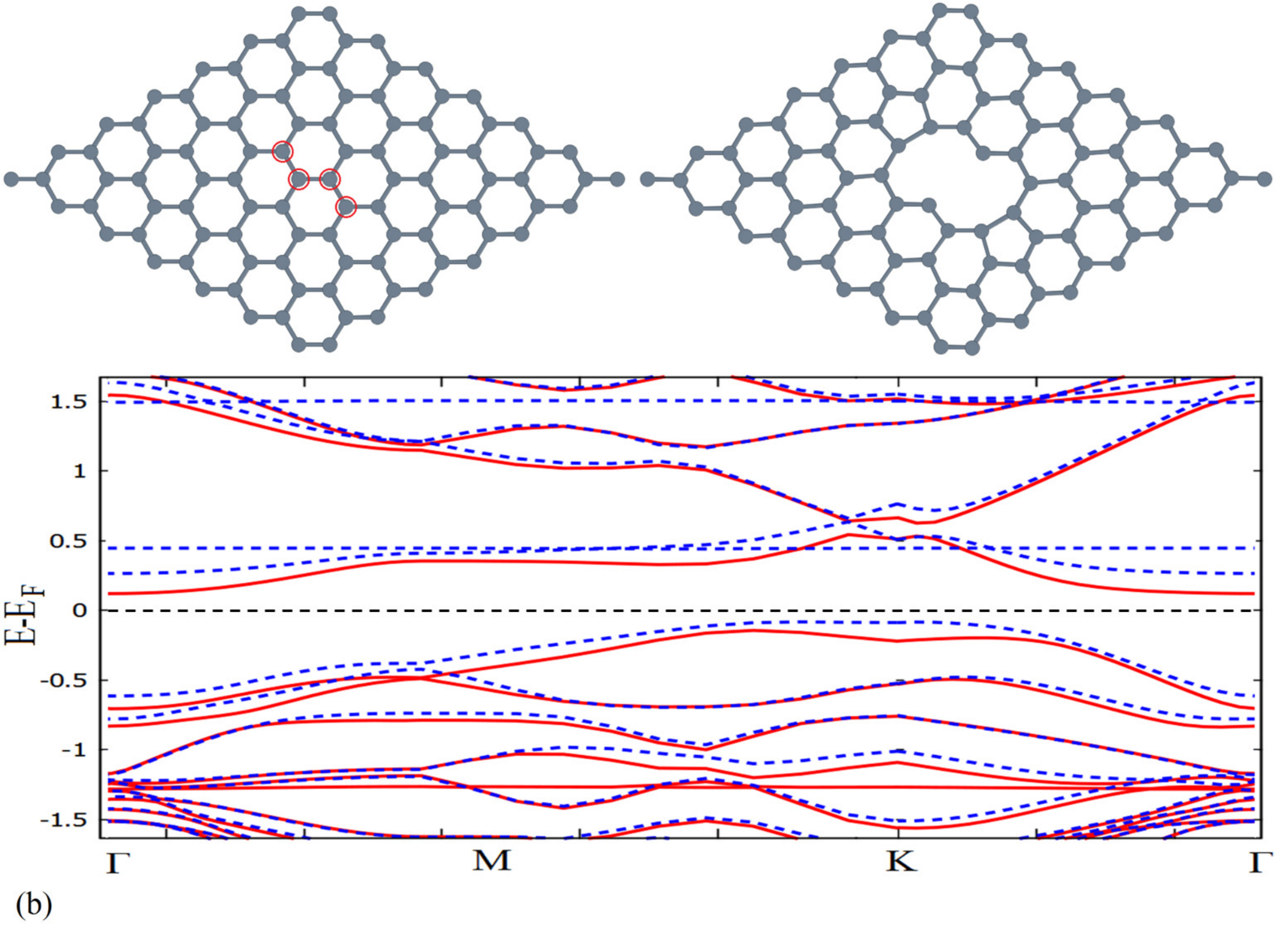}}\\
\resizebox{.45\textwidth}{!}{%
\includegraphics[width=1cm]{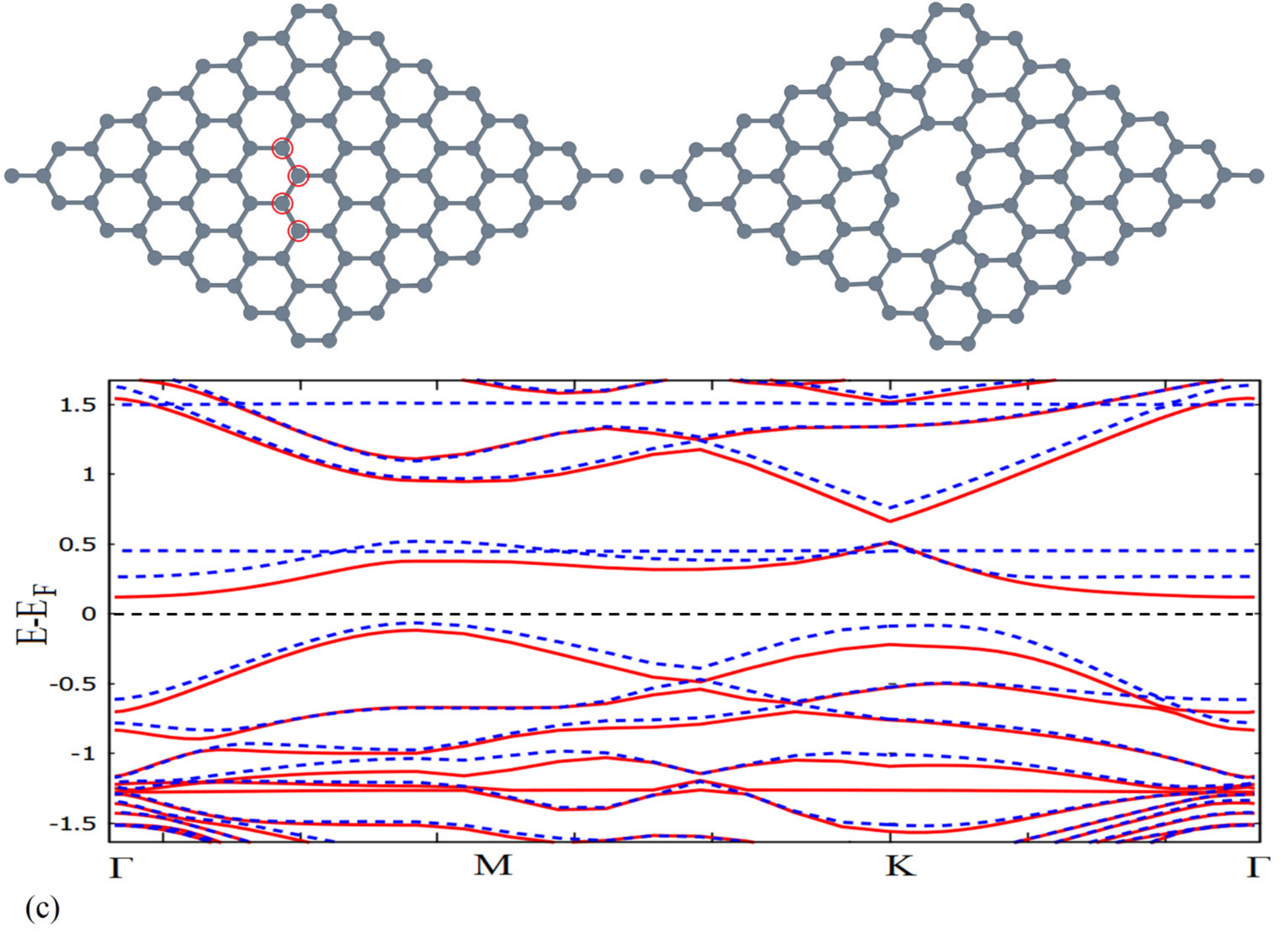}}\\
\resizebox{.45\textwidth}{!}{%
\includegraphics[width=1cm]{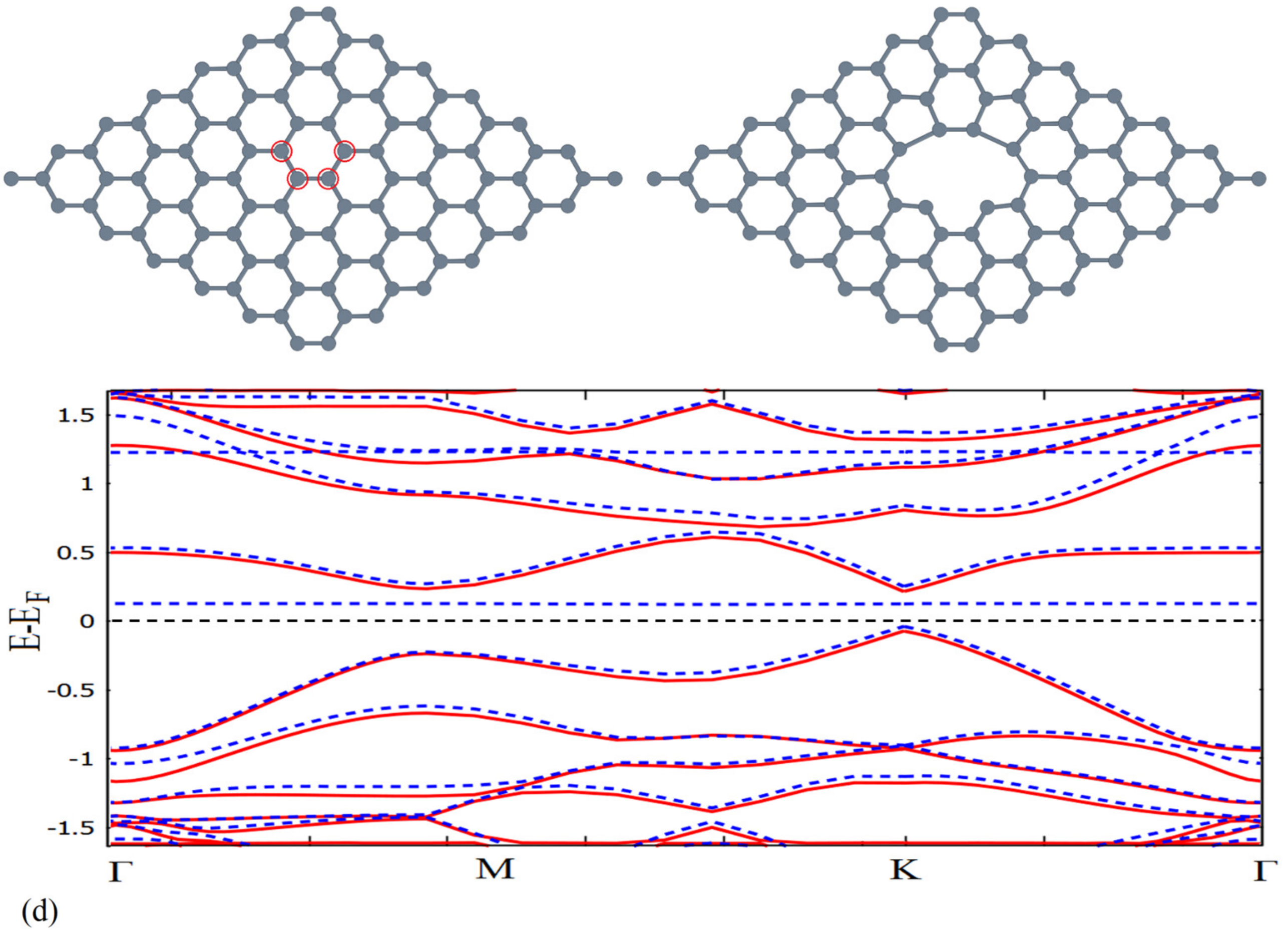}}
\end{tabular}
\caption {Different topologies of four vacancies in the supercell; their band structures, formation energies 
and the magnetic moments.}
\label{Vac4}
\end{figure}

In order to create a tri-vacancy structure we remove three nearest-neighboring carbon atoms belonging to one hexagon
in the center of the supercell. The relaxed structure, depicted in Fig. \ref{Vac1}c, creates {\it 55-10} defect consisting 
of two pentagons and one decagon. The band structure and DOS of a graphene with a tri-vacancy defect is presented in
Fig. \ref{Vac1}c. Tri-vacancy introduces four defect levels, two of which narrows back the band gap and other two levels 
cross the Fermi level, transforming the graphene to metallic state.
 Apart from the tri-vacancy defect two other structures with three vacancies are considered, which are produced
(i) by removing two nearest-neighboring carbon atoms in a hexagon but the third one is located in the next-nearest
to these two atoms as it is depicted in Fig. \ref{Vac3}a, producing three pentagon and one undecagon defect,
(ii)by removing three carbon atoms in a hexagon all belonging to one sublattice as it is shown in Fig. \ref{Vac3}b,
which produces three nonagon and one triangle. All these configurations with three vacancies contain 
dangling bonds, and therefore, reveal a magnetic moment. Nevertheless the formation energy and the magnetic 
moment of the tri-vacancy of {\it 55-10} defect take the values $8.84~eV$  and $1.00~\mu_B$, 
correspondingly (see, Table \ref{table2}). The values of the formation energy and the magnetic moment for other 
configurations {\it 999-3} and {\it 555-11} are respectively $17.49 eV$, $4.00 \mu_B$ and $11.91 eV$, $1.54 \mu_B$ 
(see, Table \ref{table2}), which are much more higher than those given for tri-vacancy {\it 55-15}.
\begin{table}[t]
\caption{The formation energy and the total spin polarization for two vacancies in different positions 
shown in Figs. \ref{Vac2}.}
\begin{tabular}{c  c  c} 
\hline \hline
\textbf{Distance} &                     \textbf{Formation energy}       & \textbf{Spin polarization}\\ 
\textbf{(in interatomic dist. d)}    &  \textbf{(eV)}                   & \textbf{$(Q_{up}-Q_{down})$} \\ [0.5ex]  
\hline
1d (DV)         & 5.76       & 0.0  \\ 
2d              & 13.027     & 2.614 \\
3d              & 13.197     & 2.003  \\
4d              & 13.547     & 2.339  \\
5d              & 13.435     & 2.016 \\ [1ex] 
\hline \hline
\end{tabular}
\label{table3}
\end{table}  
The tetra-vacancies of different configurations are depicted in Fig.\ref{Vac4}. Among of all these 
configurations only one structure {\it 555-9} containing three pentagon and one nonagon defects  has the minimal 
formation energy $7.91~eV$ and zero magnetic moment (see, Table \ref{table2}). The relaxed supercell and the 
band structure of {\it 555-9} tetra-vacancy is depicted in Fig. \ref{Vac4}a. The opposite spin-polarization bands of 
the {\it 555-9} tetra-vacancy are highly degenerated. Furthermore, the Fermi energy in this case crosses the vacancy 
level yielding an extended state. Two other tetra-vacancy configurations, the relaxed supercells and the band 
structures of which are shown in Fig. \ref{Vac4}b, and c, contain exactly the same defect structure {\it 5-12-5} 
of two pentagons and one dodagon and have the same formation energy $\sim 11.27~eV$ and the magnetic moment  
$\sim 2.02~\mu_B$. The tetra-vacancy structure {\it 55-12} of two pentagon and one dodagon defects, presented in 
Fig. \ref{Vac4}d, is produced by removing four nearest-neighboring atoms in a hexagon. The same relaxed 
supercell with {\it 55-12} defect is obtained by removing two nearest-neighboring atoms in one side and two similar 
atoms in other side of the hexagon so that these pairs are not connected each other. The formation energy and 
the magnetic moment of this type of defects are $13.33~eV$ and $1.99~\mu_B$ correspondingly.  A small band gap 
is opened around the Fermi level in these configurations, the band structures of which are presented in 
Fig. \ref{Vac4}b, c, and d, yielding quasi-localized states. Note that there is an imbalance of the vacancies 
between the sublattices A and B in the tetra-vacancy depicted in Fig. \ref{Vac4}a, since 3 vacancies belong to 
one sublattice and only one vacancy belongs to other sublattice. Nevertheless, the number of vacancies located 
in A and B sublattices are equal each other in other three tetra-vacancy configurations of Figs. \ref{Vac4}b, c, and d.
Note that tri- and tetra-vacancy have been studied \cite{tlhb14} by coalescing a mono-vacancy to
di- and tri-vacancy, correspondingly. Therefore, the obtained morphologies correspond to those
drawn in Figs. \ref{Vac4}b, c, and d with non-zero magnetic moments. 

The discussion above show that SVs in a graphene are rather mobile defects, which may migrate and coalesce 
with other single vacancy forming a DV or with other polyvacancies even at temperatures slight higher than that
of a room temperature. DVs are immobile defects at room temperature. Nevertheless they can migrate either
at higher temperatures or by means of transformation to other DV modifications. The mobile defects are reincorporated
into the crystal structure at  dislocations, grain boundaries and other defect sinks or at the crystal external surfaces.  

Summarizing all these calculation results, one can conclude that (i) among of the vacancy clusters with even
vacancies there is at least one cluster where the dangling bond is absent. The magnetic moment of such 
vacancy cluster is zero, and the formation energy of such cluster is less than those of all other clusters 
with the equal number of the vacancies; (ii) among of the clusters with odd vacancies there is at least  
one configuration which contains only one dangling bond. The formation energy of such cluster is minimal 
among of all clusters with equal number of vacancies, and the magnetic moment of this vacancy cluster
in $\sim 1~\mu_B$; (iii) the vacancy induced magnetic moment of the graphene does not obey the Lieb's
rule, which states that the magnetization of a bi-lattice structure is determined with the difference of 
the atomic number in the sub-lattices. Our calculations show that the vacancy induced magnetic moment
is determined with the number of the dangling bonds in the structure; (iv) the formation energy of $N$
vacancies, coalescenced into a single hole is less than the other aggregations of $N$ vacancies with 
several pieces. Therefore, mono-vacancies in a graphene migrate and are collected in a hole-like structure
with lowest in energy.  In a real crystal, migration of mobile vacancies takes place toward an extended 
defects existing in the structure like grain boundary, dislocation or a sample edge, segregating 
from the parent structure and healing the crystal.
 
\section{Segregation of vacancies and healing effect in graphene}
Recent transmission  electron microscopy (TEM) \cite{hsgu04, mker08, gmer09, ltjn14} and scanning tunneling microscopy 
experiments \cite{tdn08} reveal defects and vacancies and their migration in graphene with atomic resolution. 
Irradiation of a graphene sample with high energy electrons or ions creates homogeneously distributed single 
vacancies. The migration barrier of a single  vacancy in graphene was calculated by several groups 
\cite{eteh03, klfn06, tlhb14} yielding about $1.3~eV$.  We examined that a formation energy 
of divacancy is lower than that of two isolated monovacancies. This fact has been shown in other works 
\cite{lwyh05, syo07} too. Therefore, the single isolated vacancy is rather mobile, and it can migrate 
and has a tendency to coalesce either with other single vacancy to form a divacancy \cite{lwyh05} or 
with an extended defect like grain boundary, dislocation or the sample edge. The migration energy 
of a DV was calculated \cite{eteh03} to be around of $7~eV$, which is much higher than that presented 
above for a single vacancy. Nevertheless, migration of the divacancies has been recently reported 
in Girit et al. work \cite{gmer09}, where the real-time dynamics of carbon atoms were visualized in a defective
graphene using the aberration-corrected TEM technique. The subatomic 
resolution of the TEM images allows us to observe a formation of divacancies and their diffusion in the 
crystalline structure. 
In the previous Section we calculated a formation energy of three vacancies (in one hexagon) and a four 
(connected) vacancies in the supercell, which can be realized in 3 and 4 different forms, depicted in 
Figs. \ref{Vac1}, \ref{Vac3} and \ref{Vac4}, correspondingly. Among of all these configurations only 
tri-vacancy (Fig. \ref{Vac1}c) and tetra-vacancy (Fig. \ref{Vac4}a) have minimal formation energy and 
minimal magnetization, allowing the vacancies to inosculate into a big cluster. 

Irradiation produces point defects, which are distributed randomly throughout the sample. The flux of atoms and defects causes 
a buildup or depletion of alloying defects and/or vacancies in the vicinity of dislocations, grain boundaries or the 
crystal surface. 

In order to study dynamics of atoms and defects in a graphene, resulting in their redistribution, we use 
extensive kinetic model of segregation in dilute alloys \cite{jl76, jl77, jl78, wol79}. The possibility 
that impurities or alloying elements might segregate and form second phases at 
internal surfaces such as voids or on the external surfaces during irradiation was reported in many publications 
\cite{anthony72, ohl73, oh72, oswt74, sth73, ow74, ropw78, jl78, wol79} and was first confirmed experimentally in 
a high-purity $18Cr-8Ni-1Si$ stainless steel during in situ bombardment in a high-voltage electron microscope 
\cite{ohl73, oh72}, in heavy-ion bombarded vanadium \cite{oswt74, sth73}, stainless steels \cite{ow74}, and in 
nickel binary alloys \cite{ropw78}. The vacancies in the graphene layer are 
considered to be in a thermal equilibrium state during diffusion process, which imposes that the thermal equilibrium 
concentration of the vacancies is reached in a time considerably smaller than the diffusion time. A diffusion 
process is characterized by the concentration gradient of the carbon atoms or the vacancies, which is the only 
factor of the nonequilibrium process.

As a consequence of irradiation of the graphene with ions  (e.g., with fluorine) or with high-energy protons and 
carbon ($C^{4+}$) ions \cite{nstl12}, the local concentrations  of $C_A$ carbon  and $C_B$ alloying 
atoms as well as the local concentrations of $C_V$ vacancies and $C_i$  interstitials change according to the 
following continuity equations \cite{wol79,  ow74, ropw78, jl78} for atoms or defects fluxes,
\begin{eqnarray}
&&\hspace{-7mm} \frac{\partial C_A}{\partial t}=\nabla \left[D_A \alpha \nabla C_A  +S C_A\left(d_ {Ai}\nabla C_i-d_{AV} 
\nabla C_V\right)\right],
\label{carbon}\\
&&\hspace{-7mm} \frac{\partial C_V}{\partial t}=\nabla \left[-(d_{AV}-d_{BV})\alpha S C_V \nabla C_A  +D_V\nabla C_V\right] +
\nonumber\\
&&K_0-R,
\label{vacancy}\\
&& \hspace{-7mm}\frac{\partial C_i}{\partial t}=\nabla \left[(d_{Ai}-d_{Bi})\alpha S C_i \nabla C_A  +D_i \nabla C_i\right] + 
K_0-R, 
\label{interstitial}
\end{eqnarray}
where $K_0$ and $R$ are correspondingly the rates of production and recombination of vacancies and interstitials by 
irradiation, $S$ is  the average surface area of two-dimensional (2D) sample under investigation. The first term on 
the right-hand side of Eq. (\ref{carbon}) is the atom (carbon atom and adatom in our case) fluxes induced by 
the chemical-composition gradient, the second and third terms are the atom fluxes driven by the interstitial 
and vacancy gradients, respectively. The defect fluxes represented by the right-hand-side of Eqs. (\ref{vacancy}) 
and (\ref{interstitial}) for vacancies and interstitials are driven by the $A$- and $B$-atom concentration gradients 
and by their own gradients, respectively. The total diffusion coefficients for the various species $D_A$, $D_B$ or 
$D_V$, $D_i$ are written as \cite{wol79},
\begin{eqnarray}
&&D_A=d_{AV}N_V+d_{Ai}N_i,\nonumber\\
&&D_V=d_{AV}N_A+d_{BV}N_B,\label{diffusion}\\
&&D_i=d_{Ai}N_A+d_{Bi}N_B,\nonumber\\
\end{eqnarray}
where $N_A$ ($N_B$) and $N_V$ ($N_i$) are the $A$ ($B$)-atom fraction and the atomic fraction of vacancies (of interstitials),
respectively. Two terms in the expression of $D_A$ are the partial diffusivity coefficients of $A$-atoms respectively
via  $N_V$ atomic fraction of vacancies  and via $N_i$ atomic fraction of interstitials. The diffusivity coefficients $d_{AV}$
and $d_{Ai}$  are given as  \cite{comment1},
\begin{equation}
d_{AV}=\frac{1}{4}b_{V}^2 z_{V}\nu_{AV}, \quad d_{Ai}=\frac{1}{4}b_{i}^2 z_{i}\nu_{Ai},
\end{equation}
where $\nu_{AV}$ ($\nu_{Ai}$) is the jump frequency for the exchange of a given $A$-atom-vacancy ($A$-atom-interstitial)
pair, $b_V$ ($b_i$)  and $z_V$ ($z_i$) are correspondingly the jump distance and the coordination number for a vacancy
(for an interstitial).  
The similar parameters can be defined for a diffusion of $B$-atoms via interstitials and vacancies.
\begin{figure}
\resizebox{.48\textwidth}{!}{%
\includegraphics[width=1cm]{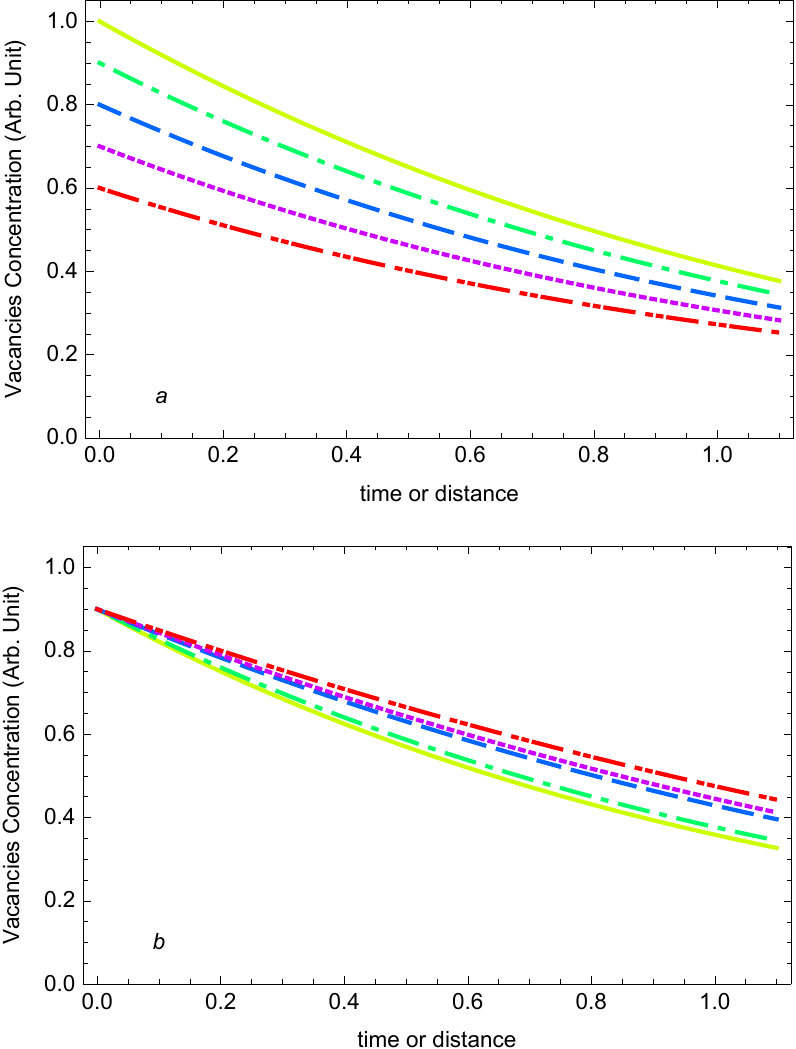}}
\caption {Distribution of the vacancy concentration with time (or in space)in a sample with dimensionless length $L=7$ 
(a) at fixed value of $\alpha =1.2$ and for $N_{0V}=1.0$ solid (yellow) curve, $N_{0V}=0.9$ dot-dashed (green) curve, 
$N_{0V}=0.8$ dashed (blue) curve, $N_{0V}=0.7$ dotted (violet) curve, and $N_{0V}=0.6$ double dot-dashed (red) curve; 
and (b) at fixed value of $N_{0V}=0.9$ and  for $\alpha=1.1$ solid (yellow) curve, $\alpha=1.2$ dot-dashed (green) curve, 
$\alpha =1.5$ dashed (blue) curve, $\alpha =1.6$ dotted (violet) curve, and $\alpha =1.8$ double dot-dashed (red) curve.}
\label{Vacancy}
\end{figure}
The equation for the element $B$ is omitted according to the relation $C_B=1-C_A$. 
The thermodynamic factor $\alpha$ in expressions (\ref{carbon})-(\ref{interstitial}) 
\begin{equation}
\alpha=1+\frac{\partial \ln \gamma_A}{\partial \ln N_A}=1+\frac{\partial \ln \gamma_B}{\partial \ln N_B},
\end{equation}
takes care of the difference between the chemical potential gradient \cite{wol79}, which is the true driving force for 
the diffusion of  $A$- and $B$-atoms, and the concentration gradient. $\gamma_A$ and $\gamma_B$ are the activity 
coefficients. The thermodynamic factor deviates from unity for non-ideal solutions.

The coupled system of equations (\ref{carbon})-(\ref{interstitial}) is a set of {\it the non-linear partial differential 
equations} in two variables, which was solved numerically in many works \cite{ohl73, oh72, oswt74, sth73,ow74, ropw78, jl78, wol79}.
In order to study dynamics of the vacancy concentration in a graphene, we simplify expressions (\ref{carbon})-
(\ref{interstitial}), written for one-dimensional case, by ignoring the interstitials ($C_i=0$) and alloying 
atoms ($C_B=0$), also by converting the surface concentrations $C$ into the atomic fractions $N$ according to the relationship 
$N_A=S C_A$ and $N_V=SC_V$. We assume that an irradiation of the system is finished, and further generation of
the vacancies and defects is stopped, $R_0=0$ and $K=0$. By turning to new dimensionless variables $\rho=x/b_V$ for
the spatial coordinate and $\tau=z_V \nu_{AV}t/4$ for the time, the equations for the dimensionless atomic fraction 
$N_A=C_A S$ and $N_V=C_V S$ read as,
\begin{eqnarray}
&&\frac{\partial N_A}{\partial \tau}=\frac{\partial}{\partial \rho}\left[\alpha N_V\frac{\partial N_A}
{\partial \rho}-N_A\frac{\partial N_V}{\partial \rho}\right],
\label{carbon1}\\
&&\frac{\partial N_V}{\partial \tau}=\frac{\partial}{\partial \rho}\left[N_A\frac{\partial N_V}
{\partial \rho}-\alpha N_V\frac{\partial N_A}{\partial \rho}\right].
\label{vacancy1}
\end{eqnarray}
Migration of a vacancy throughout the crystal is realized by hopping of a carbon atom over the vacancy. Therefore, 
the time evolutions of the atoms and the vacancies are opposite each other. 
Equations (\ref{carbon1}) and (\ref{vacancy1}) are second-order non-linear coupled equations for $N_A(\rho, \tau)$
and $N_V(\rho, \tau)$. Furthermore, the atomic fractions for carbon atoms $N_A(\rho, \tau)$ and for vacancies 
$N_V(\rho, \tau)$ are linked according to 
\begin{equation}
N_A(\rho, \tau)=1 - N_V(\rho, \tau),
\label{symmetry}
\end{equation} 
since the total concentrations of carbon atoms and vacancies at arbitrary point is conserved
in the absence of interstitials as well as of production and recombination of the vacancies under irradiation. 

We introduce new coordinate $\xi= \rho + \tau$ and demand that $N_A$ and $N_V$ depend only on $\xi$, 
so that $N_A(\rho, \tau)=N_A(\xi)$ and $N_V(\rho, \tau)=N_V(\xi)$. It is worthy to note 
that such transformation resembles the {\it soliton} transformation for the Korteweg-de Vries (KdV) equation\cite{dj89}. 
Then  Eqs. (\ref{carbon1}) and (\ref{vacancy1}) can be written under this assumption and the condition 
(\ref{symmetry}) in the following form,
\begin{eqnarray}
\hspace{-5mm}&&\frac{\partial}{\partial \xi}\left\{\alpha (1- N_A) \frac{\partial N_A(\xi)}{\partial \xi} + N_A
\frac{\partial N_A(\xi)}{\partial \xi} -N_A \right\}=0; \label{eq-CA}\\
\hspace{-5mm}&&\frac{\partial}{\partial \xi}\left\{(1- N_V) \frac{\partial N_V(\xi)}{\partial \xi} + \alpha N_V
\frac{\partial N_V(\xi)}{\partial \xi} +N_V \right\}=0,
\label{eq-CV}
\end{eqnarray}
Further we will study only the Equation (\ref{eq-CV}) for the vacancy fraction, and the solution for $N_A(\xi)$
can be determined according to Eq.(\ref{symmetry}). Integration of Eq. (\ref{eq-CV}) yields
\begin{eqnarray}
\hspace{-5mm}&&\left(1- N_V \right) \frac{\partial N_V(\xi)}{\partial \xi} + \alpha N_V(\xi)
\frac{\partial N_V(\xi)}{\partial \xi} -N_V =N_0. 
\label{eq-CV2}
\end{eqnarray}
This equation (\ref{eq-CV2}) is once more integrated yielding,
\begin{eqnarray}
&&\left[1 +(1-\alpha)N_0\right] \ln \left|\frac{N_A(\xi)+N_0}{N_0}\right| -\nonumber\\
&&(1-\alpha)\left[N_A(\xi)+N_0\right]=\xi+N_1 .
\label{vacancy2}
\end{eqnarray}
The constants $N_0$ and $N_1$ have to be determined from the boundary conditions. A 1D sample of the dimensionless 
length $L$ is assumed to be free of the internal defects and sinks, so that segregation 
could occur  only to the edge. The conditions are imposed to the half of the sample because of the symmetry 
of the problem. The vacancy is assumed to be created at $t=0$ at the center of the sample $x=0$ with concentration
of $N_V(x=0, t=0)\equiv N_{0V} = \exp(-E_V/k_BT)$, where $E_V$ is the formation energy of a vacancy. At the same 
time, the vacancy concentration and flux at the boundary $\xi=L/2$ is given as 
$N_V(\xi=L/2)=\left[1-\exp(-E_V/k_BT)\right]/L=(1-N_{0V})/L$ and $\left. \frac{\partial N_V}{\partial \xi}\right|_{\xi=L/2}=0$. 
These conditions yield, 
\begin{eqnarray}
&&N_0=\frac{N_{0V}-1}{L}=\frac{1}{L}\left\{\exp \left(-\frac{E_V}{k_BT}\right)-1\right\};
\label{N0V}\\
&&N_1=\left[1-(1-\alpha)\frac{1-N_{0V}}{L}\right] \ln \left| \frac{N_{0V}+\frac{N_{0V}-1}{L}}{\frac{N_{0V}-1}{L}+N_{0V}}\right| -\nonumber\\
&&(1-\alpha)\left(N_{0V}+\frac{N_{0V}-1}{L}\right).
\label{N1V}
\end{eqnarray}
Finally, the time- /coordinate-evolution of $N_V(\xi=x/b_V+z_V\nu_{AV}t/4)$  reads as
\begin{eqnarray}
&&\left[1 +(1-\alpha)\frac{N_{0V}-1}{L}\right]\ln \left|\frac{N_V(x, t)+\frac{N_{0V}-1}{L}}{N_{0v}+\frac{N_{0V}-1}{L}}\right|-\nonumber\\
&&(1-\alpha)\left[N_V(x, t)-N_{0V}\right]=\frac{x}{b_V}+\frac{1}{4}z_V\nu_{AV}t.
\label{vacancy3}
\end{eqnarray} 

Fig. \ref{Vacancy} shows the vacancy density dependence on time or
coordinate. We assume that the vacancy concentration of $N_{0V}(T)=e^{-\frac{E_V}{k_BT}}$ is created at $t=0$ at 
the middle of the sample $x=0$ with length $L$ by irradiation. Irradiation is stopped just at $t=0$, and the system 
is relaxed to an equilibrium state by migration of the carbon atoms over the vacancies. Fig. \ref{Vacancy}a shows 
migration of the vacancy concentration, created at $x=0$ point at $t=0$ with different concentration $C_{0V}$ 
for fixed parameter $\alpha$, with time, so that the vacancy concentration at $x=0$ decreases with time and 
segregates to the boundary of the sample.  Figure \ref{Vacancy}b displays the same dependence that is in 
Fig. \ref{Vacancy}a for fixed value of $C_{0V}(T)$ but for different values of $\alpha$.

\section{Conclusions}
Considerable attention has been paid recently to understand magnetism of a single layered graphene. In this 
work we studied influence of the vacancies on the magnetic properties and healing effects of a graphene. 
Our {\it ab-initio} DFT investigations of mono-, di-, tri- and tetra-vacancy in the single-layer graphene show 
that the formation energy of many-vacancies is lower when the vacancies coalesce together forming a big cluster 
(hole) instead of homogeneously distribution over the structure as single vacancies. Furthermore, the formation 
energy is minimal when the $\sigma$-bonds, appearing in the process of vacancy creation, rebind each other reducing
the dangling bonds to the minimal number in the internal edge of the hole.  In this case, a contribution 
to the magnetic moment yields the dangling bonds in the internal edge of the hole, the number of which is
much less than the total number of the vacancies. 

The process of clustering of the vacancies is similar to the segregation process in binary compounds. 
In a crystalline binary bulk (3D) compound the segregation may results in separation of the structure 
into two coexistent crystalline substructures or into film on the host crystal in the compound. 
Such a picture of a clustering and forming the holes under irradiation, instead of homogeneous creation 
of vacancies in a graphene, seems to be a result of two-dimensional character of the structure and absence 
of the off-diagonal long range order (ODRLO) \cite{mw66}.

ODLRO fails in the two-dimensional (2D) crystals due to the Mermin-Wagner theorem \cite{mw66, rice65}. Characterizing the order
in the crystalline structure by the order parameter $\Delta ({\bf r})$ at a point ${\bf r}$, ODLRO 
is determined by the correlator $\langle \Delta ({\bf r}) \Delta ({\bf r'}) \rangle$ which behaves
as $\langle \Delta ({\bf r}) \Delta ({\bf r'}) \rangle = C \exp \left(-\frac{\xi^{(3D)}}{|{\bf r}-{\bf r'}|^{\alpha}}\right)$
for $3D$ structures, where $C>0$ and $\alpha > 1$ are constants, $\xi^{(3D)}$ is the coherence length, and as
$\langle \Delta ({\bf r}) \Delta ({\bf r'}) \rangle =  \left(\frac{\xi^{(2D)}}{|{\bf r}-{\bf r'}|}\right)^{\beta}$
for $2D$ systems, where $\beta>1$ is a constant and $\xi^{(2D)}$ is the coherence length.
Although the correlator in $3D$ system saturates at long distance 
$\lim_{|{\bf r}-{\bf r'}| \to \infty} \langle \Delta ({\bf r}) \Delta ({\bf r'}) \rangle \to C$, it vanishes in $2D$ structures.
Since long wave-length fluctuations destroy  the  ODLRO in $2D$ systems \cite{rice65}. 
The vacancies in the graphene correlate according to power-like interaction. Therefore the vacancies form a cluster (hole)
by joining each other, which competes with the $2D$ graphene structure. The vacancy induced magnetism in the graphene is
determined by the dangling bonds on the edge of the hole. On the other hand the hole structure is energetically favorable
when the dangling bonds on the edge rebind each other reducing the magnetization to a minimal value. The graphene 
structure is destroyed by reaching the holes size to a critical value.   

A graphene structure with randomly distributed vacancies can be considered as kinetically frozen-in thermodynamically
non-equilibrium states. A clustering of the vacancies into holes in a graphene sheet can be understood as
a segregation of the structure, which means a partitioning of atomic or molecular constituents into macroscopic
regions of different compositions. In order to understand segregation of vacancies in a graphene sheet we study 
analytically dynamics of the carbon atoms and vacancies  by means of non-linear diffusion equations. Exact solution 
of these KdV-like non-linear equations shows that  the vacancies, created in the middle of the sample diffuse to the 
boundary of the sample resulting a self-healing of the graphene layer.

\section*{Acknowledgments}
The reported study was funded by the Science Development Foundation under the
President of the Republic Azerbaijan-Grant No
EIF-KETPL-2-2015-1(25)-56/01/1, and partially by Azerbaijan-JINR
collaboration.

%


\begin{thebibliography}{99}
\bibitem{comment1} The multiplayer $1/6$ in the front of $d_{AV}$ and $d_{Ai}$ in Ref. [\onlinecite{wol79}], which is valid for
$3D$ systems is replaced by $1/4$ according to Fick's diffusion coeffisient for $2D$ systems.
\end{thebibliography}
\end{document}